\documentclass[11pt,twoside]{article}
\usepackage{asp2010}

\resetcounters

\begin{document}

\title{\textsf{Astro-WISE} for KiDS survey production and quality control}
\author{Gijs Verdoes Kleijn$^1$, Jelte T. A. de Jong$^2$, E. Valentijn$^1$, K. Kuijken$^2$. For the KiDS and Astro-WISE consortiums.
\affil{$^1$ Target / OmegaCEN / Kapteyn Astronomical Institute, Groningen University}
\affil{$^2$ Leiden Observatory, Leiden University, Niels Bohrweg 2, 2333 Leiden,
The Netherlands}
}

\begin{abstract}
The Kilo Degree Survey (KiDS) is a 1500 square degree optical imaging survey with the recently commissioned OmegaCAM wide-field imager on the VLT Survey Telescope (VST). A suite of data products will be delivered to ESO and the community by the KiDS survey team. Spread over Europe, the KiDS team uses \textsf{Astro-WISE} to collaborate efficiently and pool hardware resources. In  \textsf{Astro-WISE} the team shares, calibrates and archives all survey data. The data-centric architectural design realizes a dynamic 'live archive' in which new KiDS survey products of improved quality can be shared with the team and eventually the full astronomical community in a flexible and controllable manner. 
\end{abstract}

\section{The Kilo Degree Survey}

One of the radical advances that optical astronomy has seen in recent
years is the advent of wide-field CCD-based surveys. On Paranal,
ESO has recently started operating two dedicated survey telescopes:
VISTA in the infra-red wavelength region and the VLT Survey Telescope
(VST) in the optical. The lion's share of the observing time on both
survey telescopes will be invested in a set of Public Surveys.  The
largest of the optical surveys is the Kilo-Degree Survey (KiDS), which
will image 1500 square degrees in four filters ($u$,$g$,$r$,$i$) over
a period of 3--4 years. Combined with one of the VISTA surveys,
VIKING, which will observe the same area in ZYJHK, this will provide a
sensitive, 9-band multi-colour survey.

\begin{figure}[ht]
\includegraphics[width=\textwidth]{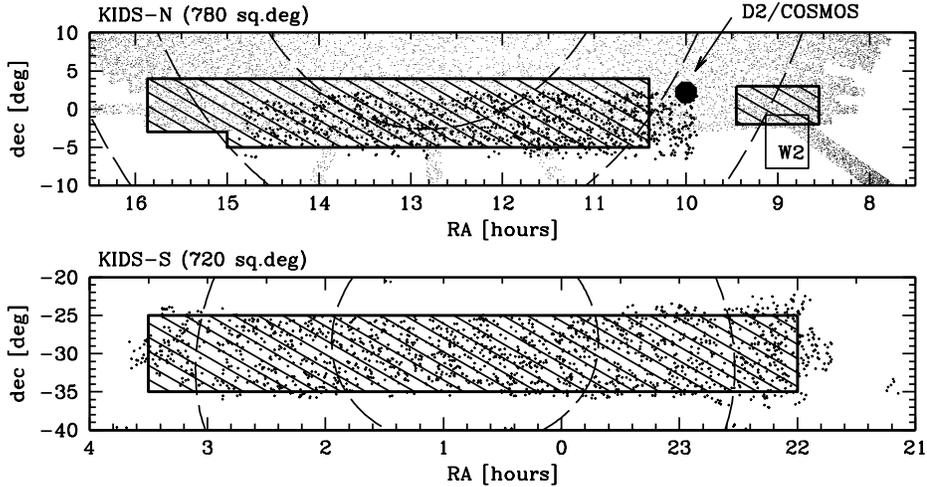}
\caption{Lay-out of the KiDS-North (top) and KIDS-South (bottom)
  fields, shown by the hatched areas. Also shown are the areas where
  2DF spectra are available, indicated by the large dots, and the area
  covered by DR7 of the SDSS survey, indicated by the small dots. The
  CFHTLS-W2 field and the DS/COSMOS deep field are overplotted on the
  top panel.}
\label{fig:areas}
\end{figure}

{\bf Observational set-up.} KiDS will cover 1500 square degrees, which is approximately 7\% of the extragalactic
sky. It consists of two patches that ensure that observations can take
place year-round. The Northern patch lies on the celestial equator,
while the Southern area straddles the South Galactic Pole
(Fig. \ref{fig:areas}). These specific areas were chosen because they
were the target of massive spectroscopic galaxy surveys already: the
2dF redshift survey \citep{2dfgrs} covers almost the same area, and
KiDS-North overlaps with the SDSS spectroscopic and imaging survey
\citep[SDSS, ][]{sdssdr8}. The exposure times for KiDS and VIKING have
been chosen to yield a median galaxy redshift of 0.8, so that the
evolution of the galaxy population and matter distribution over the
last $\sim$ half of the age of the universe can be studied. They are
also well-matched to the natural exposure times for efficient VST and
VISTA operations, and balanced over the astro-climate conditions on
Paranal (seeing and moon phase) so that all bands can be observed at
the same average rate. This strategy makes optimal use of the fact
that all observations are queue-scheduled, allowing the best seeing
time to be used for deep $r$-band exposures, for example, and the
worst seeing for $u$.

{\bf Science drivers.} The main scientific objective of KiDS and VIKING is to map the matter
distribution in the universe through weak gravitational lensing and
photometric redshift measurements. The large numbers of galaxies that
KiDS will detect, with accurate photometric redshifts up to
$z\simeq1.2$ will allow the Baryonic Accoustic Oscillations, an
important cosmological standard candle, to be measured over a large
redshift range, and thus unveil its evolution. Galaxy-galaxy lensing
(GGL) studies into the structure of galaxy halos for various redshifts
and galaxy types, will exploit the excellent image quality of the
OmegaCAM wide-field camera and the VST on the one hand and the shear
size of the KiDS data set on the other. The deep photometry and
accurate photometric redshifts also will ensure that KiDS data will be
a powerful tool to study the evolution of galaxies and clusters out to
redshifts of $z\simeq1.5$. Additionally, the extensive data set that KiDS will deliver, will be useful in a broad range of rsearch areas in astronomy, for example the study of stellar streams
and the Galactic halo.

{\bf Survey data products.} Being a Public Survey, all KiDS data will be made publicly
available. The KiDS catalogue will contain some 100,000 sources per
square degree (150 million sources over the full survey area), and for
each square degree there will be 10 GB of final image data, 15 TB for
the whole survey. A set of basic data products will be made public, both
through ESO and through the \textsf{Astro-WISE} database: calibrated coadded images, weight maps, calibration images, single-band and multi-band catalogues.
In the long-term, we intend to provide more advanced data products, for example images with
gaussianized point-spread-functions, or morphological parameters of
all detected sources.

\section{Data-centric survey handling in \textsf{Astro-WISE}}

\begin{figure}[ht]
\label{fig:astrowise}
\includegraphics[width=5.1cm]{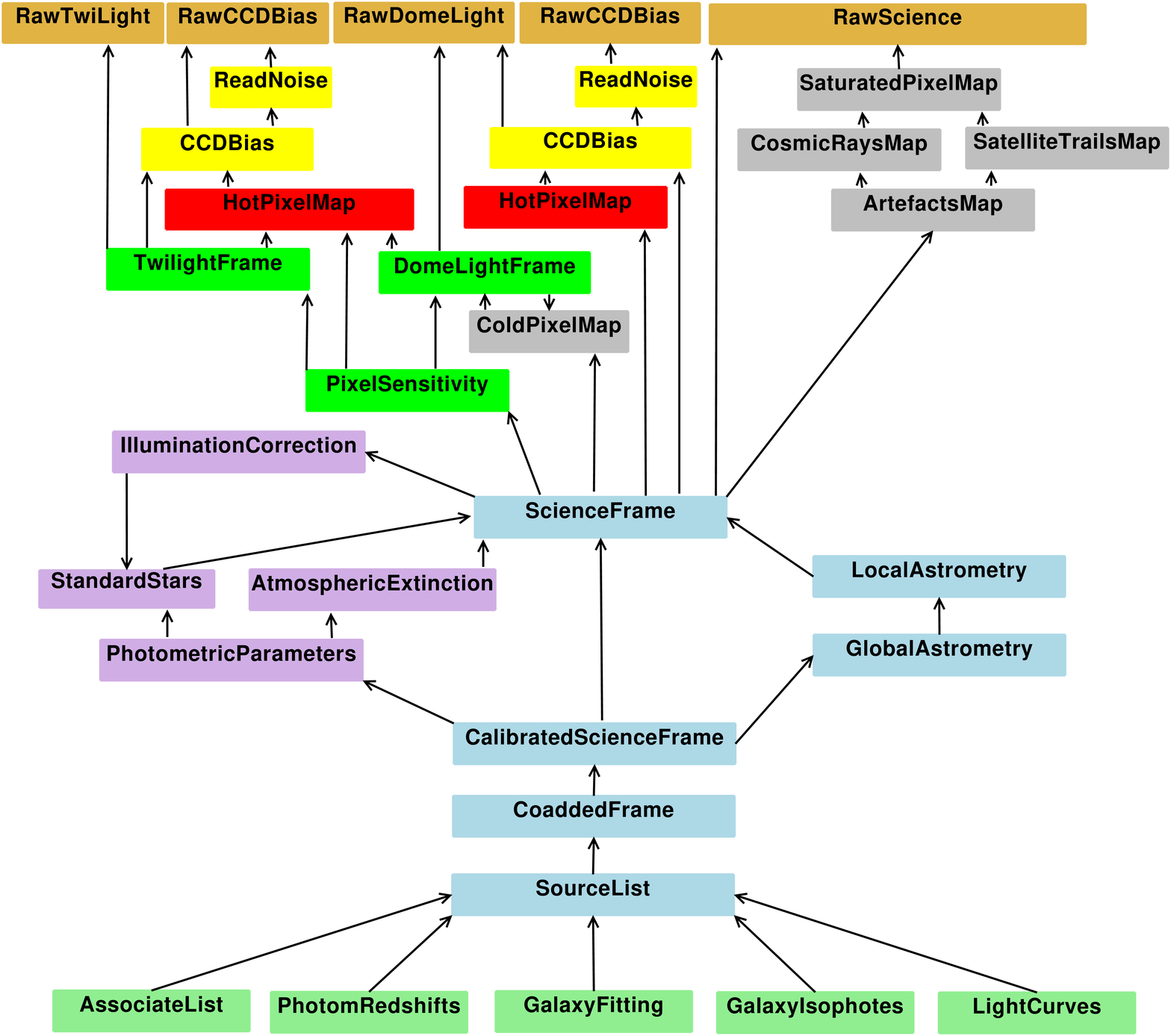}
\includegraphics[width=5.1cm]{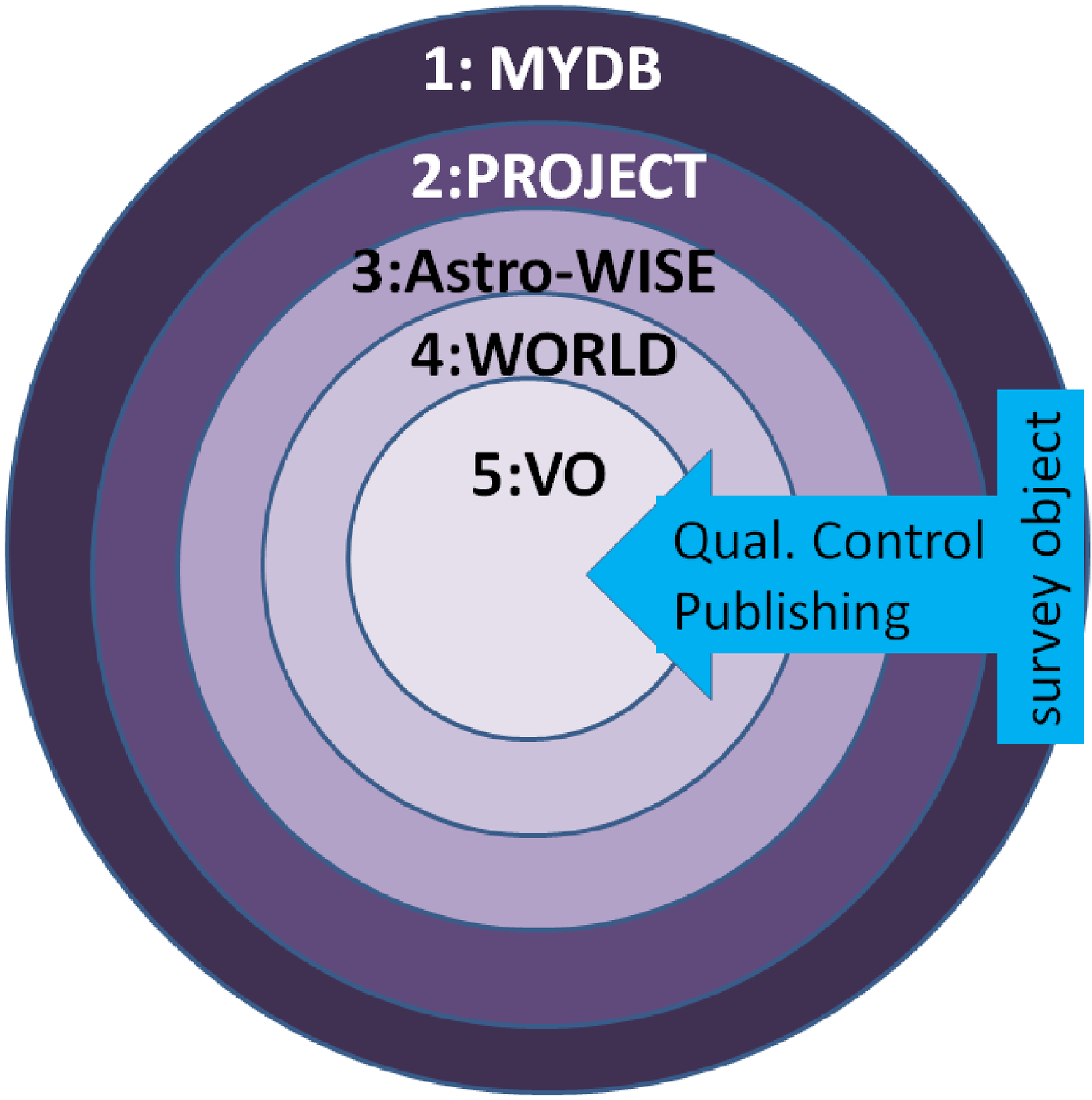}
\caption{{\bf Left:} Each box in this target diagram represents a class of survey objects. These objects not only contain the survey products denoted by familiar names in wide-field imaging. They also carry the information how they, as requested target, can be created out of other objects, illustrated by the arrows. Underlying is an object model that captures the relationship between requested information and the physics of the atmosphere-to-detector observational system.{\bf Right:} This diagram shows the survey operational levels at which data can reside within the KiDS project evironment in \textsf{Astro-WISE}.
The baseline KiDS survey products reside at 2:PROJECT. These data can be accessed only by KiDS survey team members. Each KiDS team member can experiment in her/his own level 1:MYDB to create improved versions of these baseline products. Survey data at 1:MYDB is only accessible by the single team member. If content, the member promotes the products to 2:PROJECT to share them with the team. The KiDS project manager can publish baseline survey data from 2 to levels 3 to 5. Survey data at 3:ASTRO-WISE can be accessed by all \textsf{Astro-WISE} users. At 4:WORLD, the data become accessible additionally to the astronomical community without an \textsf{Astro-WISE} account (anonymous users). At 5:VO, the data are accessible also from the Virtual Observatory.}
\end{figure}

The KiDS survey team is an international collaboration with team members at institutes spread around Europe. The European-wide hardware resources are pooled within the survey handling system \textsf{Astro-WISE} (Vriend et al., 2012, Mcfarland et al.\ 2011). In \textsf{Astro-WISE} the KiDS team members share their work on survey calibration and quality control. \textsf{Astro-WISE} is a data-centric survey handling system: all survey handling is implemented as operations by data objects on other data objects. Any type of survey product, from raw to final, is represented by a class of data objects. Survey products are framed as objects: informational entities consisting of pixel and/or metadata. Metadata is defined here as {\it all} non pixel data. The objects carrying the information of final survey products also carry the information on how they can be created out of intermediate objects. This backward chaining procedure is recursively implemented up to the raw data (see left diagram in Figure~\ref{fig:astrowise}). Thus, a request by a KiDS team member for a survey product, a target, triggers a backward information flow, in the direction of the raw data. The net effect is a forward work flow description, towards the target,  that is then executed. The backward information flow is implemented as queries to database initiated by the requested target itself. The database is queried for objects on which the target depends with the right characteristics including validity and quality. Either they exist and are returned or the query is 'backwarded' to the next level objects closer to the raw survey data. In conclusion, in \textsf{Astro-WISE} survey handling is realized by backward information flows that control forward processing steps. The information flows are the  mechanism to manage the sharing of the ocean of KiDS survey data, to control its calibration and to control and improve its quality.

{\bf Managing the survey data.} The objects of KiDS inside \textsf{Astro-WISE} are managed by having 5 different survey operational levels named privileges levels. \textsf{Astro-WISE}'s data-centric viewpoint leads to the term privileges. The object has increasing privileges to access users with numerical increase of the privileges level. The right diagram in Figure~\ref{fig:astrowise} illustrates how data is tranferred through these levels for quality control and survey delivery of the Public Survey KiDS. When using the \textsf{Astro-WISE} environment, KiDS members configure the 'Context' for their handling that includes limiting queries for and by objects to those with certain privileges.   

{\bf Survey quality control.} Objects representing survey products verify their own quality via their own verification method. It is automatically executed upon creation and sets the value of a quality flag attribute to indicate if / how its quality is compromised. Users also validate each object invoking an inspect method of the object. The user's verdict is stored with the object as a separate attribute of the object (always named is\_valid). The privileges levels serve to distinguish between experimental and baseline versions of survey data. A KiDS member tests improvements to e.g., a calibration method at the MYDB level (see Figure~\ref{fig:astrowise}). Bad outcomes are discarded by invalidating the data. Promising outcomes can be shared with the team by publishing the object to the PROJECT level (see Figure~\ref{fig:astrowise}). The fellow team members can then inspect the data and provide feedback. Upon team acceptance the object becomes baseline and can be published higher up eventually for delivery / sharing with the outside world. 
 
{\bf Survey calibration control.} Calibration data is represented also as objects in \textsf{Astro-WISE}. These objects carry a creation date and editable timestamps that mark their validity period. A request for a target generates a database query that returns all valid objects in the survey with the required validity period. The newest calibration object is then selected using the survey handling rule "newer is better". \textsf{Astro-WISE} provides webservices to manipulate this eclipsing of older calibrations by new ones by adjusting timestamps and validity. The calibration scientist uses Context to limit the survey calibration operations using these rules to the pool of calibration data at certain privileges. Calibration objects with privileges level 3 can be accessed by all \textsf{Astro-WISE} users and form a shared pool of calibration data.  

KiDS survey operations have started 15 October 2011. The KiDS team will move from a 'quick-look' versions of first survey products towards publishing of the complete KiDS Public Survey, using \textsf{Astro-WISE} as a 'live archive' that captures the accumulation of knowledge about OmegaCAM, VST and the KiDs survey data.

\begin{acknowledgements}
This work is financially supported by the Netherlands Research School for Astronomy (NOVA) and Target (www.rug.nl/target). Target is supported by Samenwerkingsverband Noord Nederland, European fund for regional development, Dutch Ministry of economic affairs, Pieken in de Delta, Provinces of Groningen and Drenthe. Target operates under the auspices of Sensor Universe.
\end{acknowledgements}

\end{document}